\documentclass{article}
\usepackage[utf8]{inputenc}

\usepackage{geometry}
\usepackage[square,sort,comma,numbers]{natbib}
\usepackage{amsmath}
\usepackage{latexsym}
\usepackage{amsfonts}
\usepackage{mathrsfs}
\usepackage{upgreek}
\usepackage{graphics}
\usepackage{graphicx}
\usepackage{epstopdf}
\usepackage{epsfig}
\usepackage{amsbsy}
\usepackage{amsfonts}
\usepackage{amsmath}
\usepackage{amssymb}
\usepackage{bm}
\usepackage{color}
\usepackage{setspace}
\usepackage{authblk}

\title{StageOpt technical write-up}

\author{Egor Dontsov and Mark McClure}
    
\begin{document}

\maketitle

\begin{abstract}
\noindent This document provides technical background for StageOpt. This tool helps to better understand fluid and proppant distribution between perforations and clusters within a fracturing stage. Limited entry technique is often used to achieve uniform fluid distribution between clusters~\cite{Cramer1987,Weddle2018,Lorwon2020}. This, however, does not automatically guarantee uniform proppant distribution. In view of this observation, StageOpt allows users to overcome this obstacle and to better design perforation design by focusing on both fluid and proppant distribution between clusters. The algorithm in StageOpt is based primarily on the paper~\cite{Dont2023c}, but it also incorporates practically important features, such as the effect of stress shadow from previous stage, perforation erosion, and other phenomena. In this write-up, flow distribution between perforations is described first, and the effects of perforation friction, near-wellbore pressure drop, as well as perforation break down and stress shadow are included. Then, proppant distribution between perforations is discussed as a function of perforation orientation. Finally, perforation erosion is added to the model. The erosion changes perforation diameter as a function of several parameters, including the flow rate through the perforation as well as the amount of proppant that flows through it. The change of perforation size in turn changes magnitude of perforation friction, which then changes the distribution of slurry, that consequently changes proppant flow distribution. Such a coupled system often leads to complex results that are hard to deduce without systematic physical modeling, which is exactly what StageOpt provides.
\end{abstract}

\section{Slurry flow distribution between perforations}

A full injection schedule is the first input to StageOpt. Each line of the schedule characterizes fluid and proppant that are pumped, as well as the injection rate and the duration of pumping. A time stepping is employed to capture the effects of varying injection parameters as well as to track the evolution of perforation erosion. A default time step is calculated by dividing the total injection time by the number of specified time steps. If the duration of a given injection period is less than the default time step, then the actual time step is reduced to match the injection period. A similar time step truncation is used to ensure that a single time step does not span over two different injection periods in the pumping schedule.

\begin{figure}
\centering\includegraphics[width=0.75\linewidth]{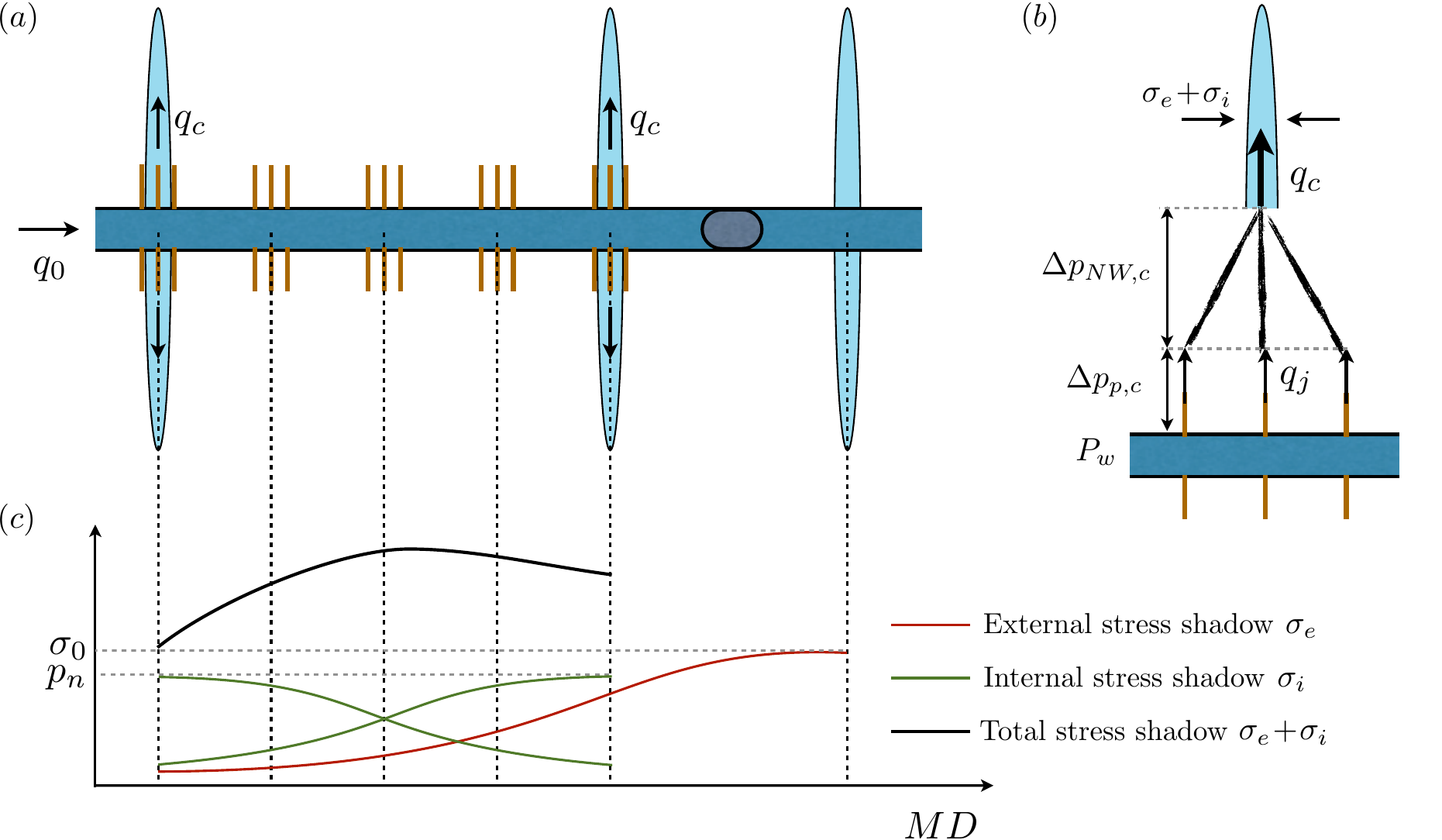}
\caption{Panel $(a)$: Schematic of initiated fractures from the stage together with the external fracture. Panel $(b)$: Distribution of pressure drops from the wellbore to the fracture for a given cluster. Panel $(c)$: Example stress shadow calculation from both internal and external fractures.}
\label{fig1}
\end{figure}

As an illustration, Fig.~\ref{fig1} shows schematic of the problem, in which only two outer fractures are initiated and there is an external fracture as well. Panel $(a)$ shows the fracture configuration, panel $(b)$ shows the flows at the cluster level, and panel $(c)$ illustrates how the total stress shadow is evaluated. Typically, outer fractures are initiated first and they create a higher stress shadow in the middle of the stage. If the perforation friction drop is unable to overcome the elevated stress shadow, then middle fractures fail to initiate. The panel $(b)$ conceptually shows pressure drops from the wellbore, across the perforations, and then across the near-wellbore zone, and finally the pressure must overcome the total stress shadow in order to have a non-trivial flow rate through a perforation.

Each perforation is assigned a value of the effective tensile strength, which is prescribed by a constant base value and uncertainty. As soon as pressure in the wellbore exceeds this value and the compressive stress, the perforation initiates and is able to take fluid and proppant. In StageOpt, one perforation is initiated at a time. However many perforations can initiate within the same time step. For instance, once the first perforation is initiated, then slurry flow is calculated and if the pressure in the wellbore is still sufficiently high, another perforation is initiated. When multiple perforations are able to initiate, the one with the smallest tensile strength is always chosen. The procedure is repeated until pressure in the wellbore becomes insufficient to initiate the remaining perforations or until all perforations are initiated. Mathematically, the initiation condition can be written as
\begin{equation}\label{initiation}
P_w > T +\sigma_e+\sigma_i+\sigma_h,
\end{equation}
where $P_w$ is the pressure in the wellbore, $T$ is tensile strength for the given perforation, $\sigma_e$ is the value of the external stress shadow from previous stage, $\sigma_i$ is the value of the internal stress shadow that is generated by the fractures that are initiated within the stage, and $\sigma_h$ is the value of the minimum horizontal stress.

The external stress shadow is assigned on the per cluster basis and all perforations within the cluster are assumed to have the same value of $\sigma_e$. In StageOpt, it is possible to populate values automatically by prescribing only the magnitude of stress shadow. In this case, the table is populated using the following variation of stress with distance: 
\begin{equation}\label{stressPKN}
\sigma_e = \sigma_0\Bigl(1-\dfrac{x_j^3}{(x_j^2+H^2/4)^{3/2}}\Bigr).
\end{equation}
Here $\sigma_0$ is the prescribed magnitude of stress shadow, $H=200$~ft is default value of fracture height, while $x$ is the horizontal distance from the external fracture to the given cluster. It is assumed that the external fracture is 30~ft away from the end of the stage. The equation~(\ref{stressPKN}) applies to constant height fractures, see e.g.~\cite{Wu2015b}. 

The values of the internal stress shadow $\sigma_i$ are calculated using the same equation~(\ref{stressPKN}), where except the magnitude of the stress shadow $\sigma_0$, the specified value of fracture net pressure $p_n$ is used, i.e.
\begin{equation}\label{stressPKN2}
\sigma_i = p_n\Bigl(1-\dfrac{x^3}{(x^2+H^2/4)^{3/2}}\Bigr).
\end{equation}
Note that the same fracture heigh is used and that each initiated fracture creates a stress shadow with the magnitude $p_n$, so the total internal stress shadow is actually a summation from all initiated fractures. Also, $x$ now is the distance from the initiated fracture to an evaluation point which represents the location of another cluster. To account for smaller magnitude of stress shadow for newly initiated fractures, the specified net pressure is scaled by the factor $V_f/V_0$. Here $V_f$ is the total fracture volume or the total volume of slurry that entered the fracture through all perforations within a cluster. At the same time $V_0=100$~bbls is the reference fracture volume. Once the fracture volume exceeds the reference volume, the stress shadow remains constant.

Once perforation initiation procedure is completed for the given time step, then slurry distribution between perforations is calculated. Perforation friction is calculated for each perforation, while the stress shadow (both internal and external) as well as near-wellbore pressure drop are calculated on the per cluster basis. Therefore, it is first assumed that the value of perforation friction is the same for all perforations within a given cluster. Let $q_c$ be the total flow rate through a given cluster that has $N_c$ perforations, then the perforation pressure drop is calculated as
\begin{equation}\label{perfpdrop}
 \sum_{j=1}^{N_c} q_j = q_c,\qquad \Delta p_{p,c} = \dfrac{\rho_f q_j^2}{2C_j^2A_j^2}.
\end{equation}
Here $q_j$ is the flow rate through $j$th perforation in the cluster, $\rho_f$ is the fluid density, $C_j$ is the discharge coefficient, while $A_j$ is perforation area. Note that the perforation pressure drop in the above equation is written in SI units. Equation~(\ref{perfpdrop}) can be solved to obtain
\begin{equation}\label{perfpdropsol}
q_j = \dfrac{C_jA_j}{\sum_j C_jA_j}q_c,\qquad \Delta p_{p,c} = \dfrac{\rho_f q_c^2}{2(\sum_j C_jA_j)^2}.
\end{equation}
The above equation provides the slurry distribution between individual perforations within a cluster for the given total cluster flow rate, as well as it calculates the perforation friction pressure drop for the whole cluster as a function of the total cluster flow rate.

Equipped with analytical solution at the cluster level~(\ref{perfpdropsol}), solution at the stage level is calculated numerically by solving the following equation:
\begin{equation}\label{floweq}
P_w = \Delta p_{p,c} + \Delta p_{NW,c} + \sigma_{e,c} + \sigma_{i,c}+\sigma_h,\qquad \sum_{c=1}^{N} q_c = q_0.
\end{equation}
Here $P_w$ is wellbore pressure, the same as used in~(\ref{initiation}), $\Delta p_{p,c}$ is perforation friction drop for the given cluster that is calculated using~(\ref{perfpdropsol}), $\sigma_{e,c}$ is the external stress shadow, $\sigma_{i,c}$ is the internal stress shadow, $q_0$ is the total injection rate, $N$ is the number of clusters, while $\Delta p_{NW,c}$ is the near-wellbore pressure drop that is calculated as
\begin{equation}\label{pNW}
\Delta p_{NW,c} = A_c q_c^{n_c}.
\end{equation}
In the above equation, $A_c$ is near-wellbore pressure drop coefficient, while $n_c$ is near-wellbore pressure drop exponent for the given cluster. In StageOpt, equation~(\ref{floweq}) is solved numerically to find pressure in the wellbore $P_w$ as well as the flow rate for each cluster $q_c$. Then, equation~(\ref{perfpdropsol}) is used to calculate slurry flow rate for each perforation.

It should also be noted that the summation in the above equations is performed only over initiated perforations that are not plugged with proppant. Otherwise, the perforation is assumed to be inactive and is removed from calculations. 

\section{Proppant flow distribution between perforations}

Proppant distribution between perforations in StageOpt depends on two primary physical phenomena: proppant settling in the wellbore and proppant turning efficiency. As the flow rate in the wellbore decreases from heel to toe, the ability of fluid to suspend particles reduces. Thus, toe clusters are typically more affected by particle settling than the heel clusters. Particle turning can be characterized by turning efficiency, which quantifies the fraction of particles that enters the perforation relative the amount of particles flowing in the wellbore in the vicinity of perforation. This parameter primarily depends on particle size and for field cases it lies within 0.7 to 0.9 range depending on problem parameters. Both particle settling and turning are described in detail in~\cite{Dont2023c}. The model is calibrated against numerous laboratory scale, field scale, as well as computational experiments~\cite{Grues1982,Nasr1989,GilliesPhD,NgameniMS,Ngameni2017,Ahmad2019a,Ahmad2019b,AhmadPhD,Ahmad2021,WuPhD,Wu2016,Wu2019,Benish2022,Snider2022,Liu2021,Wang2022}. The purpose of this section is to briefly summarize the main results.

\begin{figure}
\centering\includegraphics[width=0.75\linewidth]{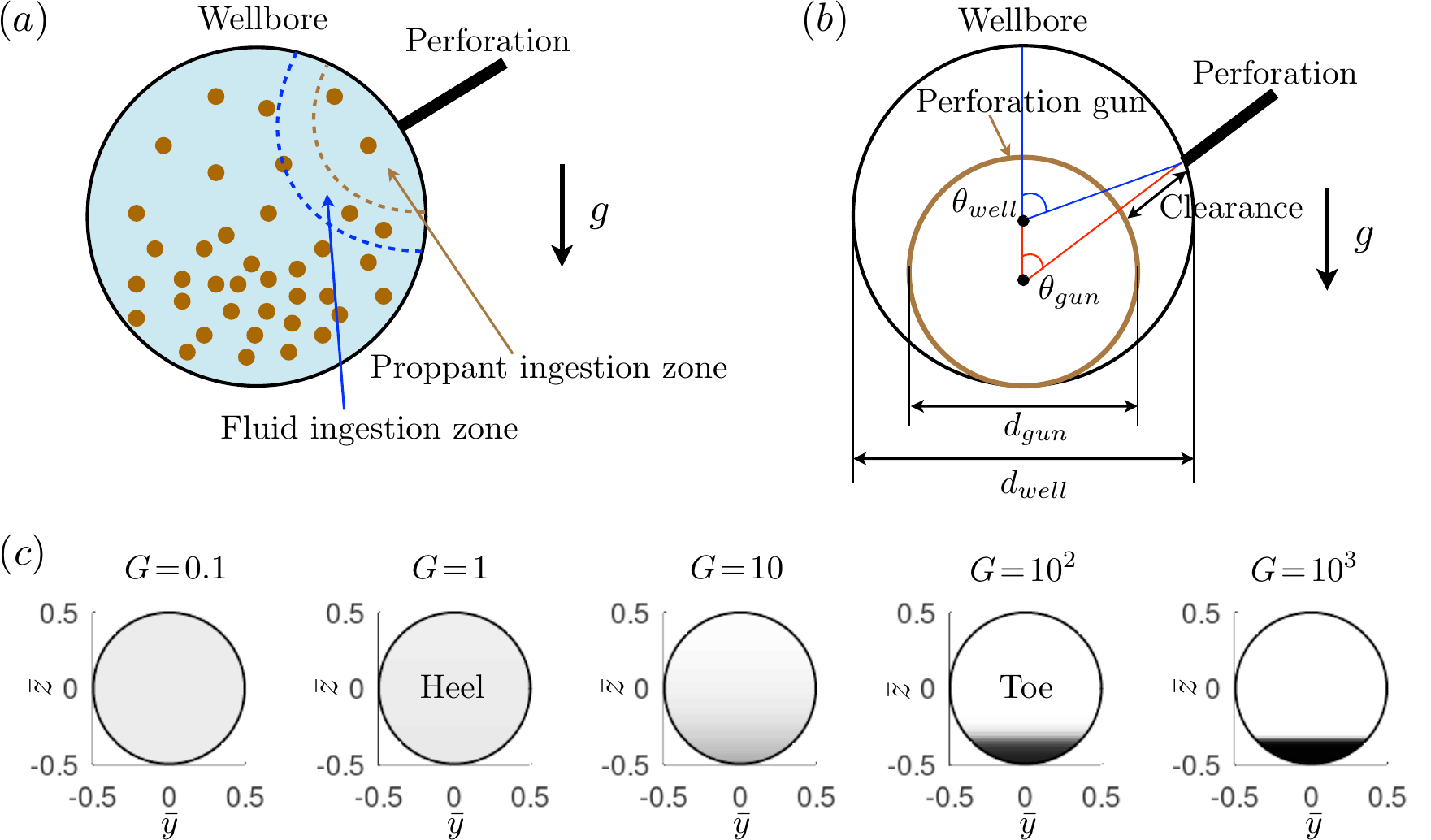}
\caption{Panel $(a)$: slurry flow in the wellbore cross-section with particle settling. Fluid and proppant ingestion zones are shown by the dashed lines. Panel $(b)$: schematic for wellbore versus perforation gun centered azimuths. Panel $(c)$: particle concentration in the wellbore for different values of the dimensionless gravity $G$, according to the model in~\cite{Dont2023c}. Figure reproduced from~\cite{Donts2023b}.}
\label{fig2}
\end{figure}

Fig.~\ref{fig2}$(a)$ shows a wellbore cross-section and schematically shows non-uniform particle distribution due to settling. Fluid and proppant ingestion zones are shown. In the ideal case of uniform velocity distribution within the cross-section, the area of the fluid ingestion zone (outlined by the dashed blue line) relative to the area of the wellbore's cross-section is equal to the slurry flow rate through the perforation divided by the total flow rate through the wellbore (upstream from the perforation). Calculations become somewhat more involved when true non-uniform velocity profile is considered, see~\cite{Dont2023c}. For typical parameters, the aforementioned fraction is often small. For instance, for the first or heel-most perforation, this ratio can be 3/90 for the case when the total injection rate is 90 bpm and each perforation receives 3 bpm. In the middle of the stage, this ratio becomes approximately 3/45. While, assuming 10 clusters and 3 perforations per cluster, immediately before the last cluster this ratio is 3/9. This clearly demonstrates that for most of the perforations, the fluid ingestion zone is relatively small or local to the perforation location. The ingestion area becomes comparable to the wellbore cross-section area only for toe perforations. The area associated with proppant ingestion is smaller than that for fluid by the multiplier of turning efficiency, so it is approximately 70\% to 90\% of the fluid ingestion area. Thus, the effect of perforation phasing becomes crucial when there is a substantial particle settling since the amount of proppant that enters the perforation depends on the value of local particle concentration near the perforation itself.

To further quantify particle settling, Fig.~\ref{fig2}$(c)$ shows particle volume fraction distribution in the well predicted by the model~\cite{Dont2023c}. The parameter that quantifies particle distribution is called dimensionless gravity:
\begin{equation}\label{Gdef}
 G = \dfrac{8\phi_m(\rho_p\!-\!\rho_f)g \cos(\theta_w)d_w}{f_D\rho_f v_w^2},
 \end{equation}
where $\phi_m=0.585$ is the maximum flowing volume fraction of particles, $\rho_p$ is particle mass density, $\rho_f$ is fluid mass density, $g=9.8$~m/s$^2$ is gravitational constant, $d_w$ is wellbore diameter, $f_D=0.04$ is a fitting parameter that can also be interpreted as a friction factor in the pipe, and $v_w$ is the average wellbore velocity or the total flow rate divided by the cross-sectional area. To account for dipping wells, the dip angle $\theta_w$ is introduced. For horizontal wells $\theta_w=0$. The only parameter that changes along the stage is the average wellbore velocity. For heel perforations the velocity is high and the value of the parameter $G$ is low (on the order of 1 for typical parameters). This leads to nearly uniform proppant distribution, as shown in the figure. At the same time, velocity is much lower for toe perforations and the dimensionless gravity $G$ can reach $100$s. This results in significantly asymmetric particle distribution, in which particle flow resembles the ``flowing bed'' state, see Fig.~\ref{fig2}$(c)$.

While the combination of particle settling and turning calculations allows to quantify the amount of proppant that enters each perforation, there are some other issues that need to be considered in the model to be practical. One such thing is illustrated in Fig.~\ref{fig2}$(b)$, which shows that the perforation gun diameter is often noticeably smaller than the inner wellbore diameter. Therefore, the azimuth relative to the center of the well $\theta_{well}$ and the corresponding azimuth relative to the center of the gun $\theta_{gun}$ are different. In the field, the azimuth relative to the center of the gun is specified, while the azimuth relative to the wellbore determines proppant distribution. If the ratio between the perforation gun and wellbore diameters is given, the different phasing definitions can be converted as
\begin{equation}\label{azimuthconversion}
\theta_{well} = \theta_{gun}+\sin^{-1}\bigl(r\sin(\theta_{gun})\bigr),\quad \theta_{gun} = \tan^{-1}\bigl((r\!+\!\cos(\theta_{well}))^{-1}\sin(\theta_{well})\bigr),\quad r = 1-\dfrac{d_{gun}}{d_{well}}.
\end{equation}
Note that care must be taken in the above relations to ensure that the angle falls into the correct quadrant.

\begin{figure}
\centering\includegraphics[width=0.9\linewidth]{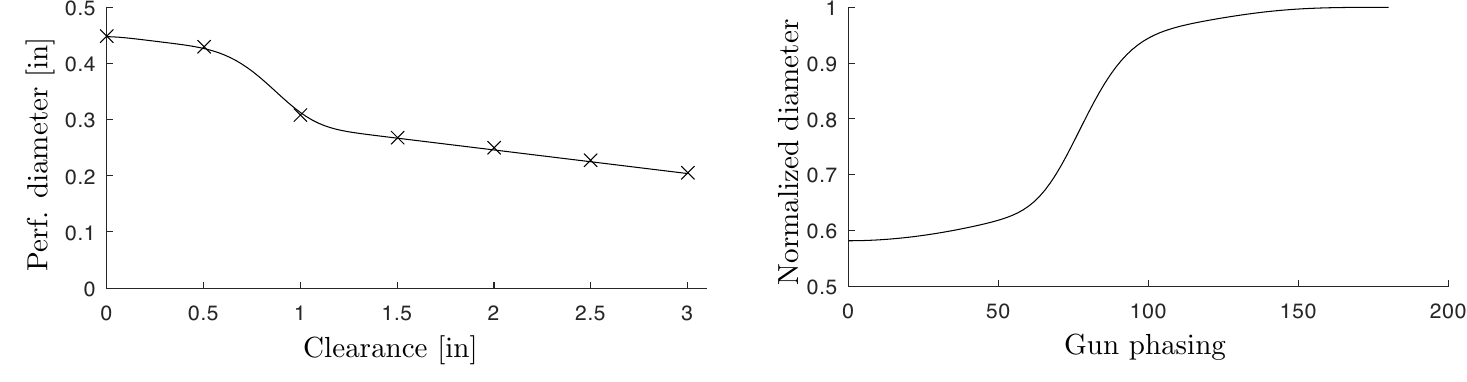}
\caption{Left: Perforation diameter versus clearance, data taken from~\cite{Bell2000}. Right: Calculated normalized diameter versus gun phasing.}
\label{fig3}
\end{figure}

In practice, perforation diameter depends on the perforation gun clearance, defined in Fig.~\ref{fig2}$(b)$. The bigger the clearance, the smaller the resultant hole diameter. While the results can vary for different shape charges and casing types, we employ data from~\cite{Bell2000} that provides a reasonable guideline. Fig.~\ref{fig3} shows the variation of perforation diameter versus clearance or the distance between the outer side of the gun and the inner side of the wellbore. The right panel in Fig.~\ref{fig3} shows the normalized perforation diameter versus phasing that is calculated from the center of the gun.
In order to calculate this dependence, the data is first fitted by a model curve (solid line on the left panel). In the calculations of the normalized diameter shown in the figure, it is also assumed that the casing inner diameter is $4.78$~in, while the gun diameter is $3.12$~in. However, in actual calculations, this assumption is not used. In other words, the result shown on the right panel in Fig.~\ref{fig3} is not universal and depends on the ratio between the gun and wellbore diameters. In StageOpt, users specify the design diameter. The actual diameter is then calculated by multiplying the design value by the normalized diameter, which is shown in Fig.~\ref{fig3}. In the simplest case, the latter normalized diameter is equal to one at the bottom of the well or when phasing is equal to $180^\circ$. For other perforation orientations, the normalized diameter is less than one. Also, StageOpt provides an ability to specify non-zero gun clearance for the design diameter. In this case, the normalized diameter is equal to one for this specified clearance. It can exceed one if the clearance is smaller and be less then one if the clearance is larger. Finally, StageOpt provides an option to specify a multiplier that linearly interpolates the resultant perforation diameter between the correlation presented in Fig.~\ref{fig3} and the specified diameter.

Field observations with downhole camera demonstrate that some perforations are screened out by proppant at the end of the treatment. To model such a phenomenon, a simple model is adopted. The probability of perforation screen-out is proportional to the total mass of proppant that flows through the given perforation. The reference probability of plugging is 5\%, while the reference proppant mass is taken as $10^4$ lbs. That is, there is 5\% chance of plugging the perforation if $10^4$ lbs of proppant is flown through it. It is possible to also specify the multiplier to change the resultant probability of screening out. In addition, a plugging bias is introduced to account for the fact that perforations that are located at the bottom of the well have a higher chance of being plugged. Note that plugging can also occur during wellbore clean-out, therefore it is important to be critical about interpreting the results. The default value of the bias is 2, but the value can be adjusted using advanced settings as:
\\
\\
Name:\\
pluggingbias\\
Value(s):\\
0\\
\\
In this example, there is no bias.

In some situations, it is desired to include the effect of proppant volume fraction on screen-out. In current implementation, it is possible to include the following multiplier for plugging probability: $(\phi/\phi_0)^n$. Here $\phi$ is the volume fraction of proppant in the slurry flowing through the perforation, $\phi_0=0.1$ is the reference value for the volume fraction, while $n$ is the screen-out exponent. Its default value is zero, so that there is no effect by default. To specify the non-zero value for $n$, users can type the following to the advanced settings:
\\
\\
Name:\\
screenoutexponent\\
Value(s):\\
1.2\\
\\
This will set $n=1.2$ for calculations. In this case, perforations that take higher proppant volume fractions have a higher tendency to get plugged. Note that once a perforation is plugged, it cannot be cleared. If all perforations are plugged, then calculation aborts.

Finally, there is an additional parameter called ``approach length''. In StageOpt, it is assumed that particles are initially perfectly suspended, and then the suspension flows the distance that is equal to the approach length before it encounters the perforated stage. If the injection rate is not sufficiently high, the particles can settle and reach a new equilibrium by the time they reach the first cluster. Default value of 100~ft is used, which is typically sufficient for field scale models. The horizontal part of the wellbore can exceed this number, but it is not that important since the goal is to ensure that the solution reaches equilibrium, which typically happens within 100~ft interval. At the same time, this setting can be crucial to design laboratory experiment with StageOpt. Then, the effect of the approach length can be much more important and there is a necessity to specify a correct value. This can be done by typing the following to the advanced settings
\\
\\
Name:\\
approachlength\\
Value(s):\\
3\\
\\
This will set the approach length to 3 ft.

\section{Perforation erosion model with lateral velocity}

Perforation erosion is crucial during hydraulic fracturing since it significantly affects fluid distribution between perforations. In standard approach, it is assumed that the degree of erosion is proportional to the proppant mass flowing through the perforation~\cite{Cramer1987}. However, in the newer approach, the erosion rate is proportional to particle energy or squared velocity of particles flowing through perforation~\cite{Long2015}
\begin{equation}\label{longerosion}
\dfrac{dD}{dt}  = \alpha C v_p^2,\qquad \dfrac{dC_d}{dt} = \beta C v_p^2 \Bigl(1-\dfrac{C_d}{C_{d,max}}\Bigr).
\end{equation}
Here $D$ is perforation diameter, $C$ is proppant concentration, $v_p$ is the average velocity of flow through the perforation, $C_d$ is discharge coefficient, $C_{d,max}$ is the maximum allowed value of the discharge coefficient, while $\alpha$ and $\beta$ are two parameters. In this model, the discharge coefficient captures the change of the shape of the perforation during the erosion, i.e. the process of transforming it to a more conical shape that is characterized by the entrance and exit diameters. Here $D$ is the smallest or exit diameter.

\begin{figure}
\centering\includegraphics[width=0.5\linewidth]{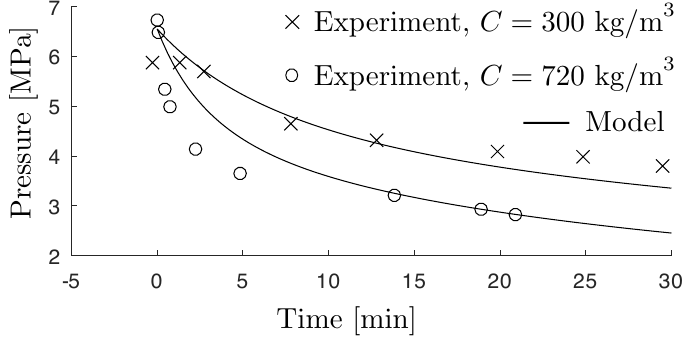}
\caption{Comparison between perforation pressure drop observed in~\cite{Crump1988} and the model, characterized by~(\ref{longerosion}) and~(\ref{ouralpha}).}
\label{fig4}
\end{figure}

The values of $\alpha$ and $\beta$ are calibrated in~\cite{Long2015} based on the experimental results presented in~\cite{Crump1988}
\begin{equation}\label{longalpha}
\alpha = 1.1\times 10^{-13}~\text{m$^2$$\cdot$s/kg},\qquad \beta = 7.47\times 10^{-9}~\text{m$\cdot$s/kg}.
\end{equation}
One issue is that the data was plotted on a logarithmic scale in~\cite{Crump1988}, but it appears that the authors in~\cite{Long2015} assumed a linear scale. This resulted in incorrect numeric values for the parameters $\alpha$ and $\beta$. With the corrected interpretation of the figure scale, the new values of the erosion parameters become
\begin{equation}\label{ouralpha}
\alpha = 3\times 10^{-13}~\text{m$^2$$\cdot$s/kg},\qquad \beta =1\times 10^{-9}~\text{m$\cdot$s/kg}.
\end{equation}
Fig.~\ref{fig4} shows the comparison of the measured perforation pressure drop versus time for two proppant concentrations, according to the results~\cite{Crump1988}. The solid lines plot the modeled value, calculated according to~(\ref{perfpdrop}). Here the perforation diameter and discharge coefficient are eroded according to~(\ref{longerosion}) and the erosion parameters~(\ref{ouralpha}) are used.
Even though the actual values of $\alpha$ and $\beta$ are different, this is not a major issue because various wellbore casings and proppants are used in practical applications. This in turn affects the actual rate of erosion. Clearly, sharper proppant and thinner casing lead to more erosion, while the combination of more round proppant and thicker casing leads to less erosion. As a result, in StageOpt the values of the perforation parameters are allowed to be changed by specifying the perforation erosion multipliers.

\begin{figure}
\centering\includegraphics[width=0.4\linewidth]{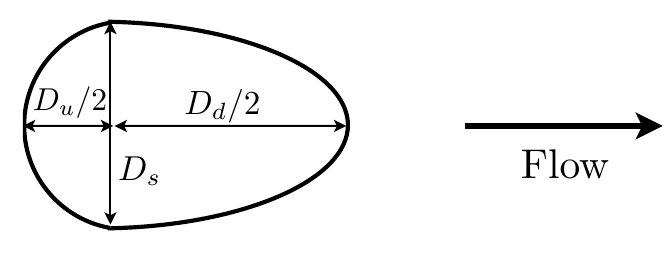}
\caption{Schematic of the eroded perforation in the model.}
\label{fig6}
\end{figure}

A interesting observation is made in~\cite{Cramer2020}. It is shown that the erosion is non-uniform between the perforations. The heel perforations erode more and also they erode predominantly in the downstream direction of the fluid flow.
To model such a behavior, the perforation erosion model~(\ref{longerosion}) is modified as
\begin{eqnarray}\label{newerosion}
\dfrac{dD_u}{dt}  &=& \alpha C v_p^2,\notag\\
 \dfrac{dD_s}{dt}  &=& \alpha C v_p^2,\notag\\
\dfrac{dD_d}{dt}  &=& \alpha C \bigl(v_p^2+\gamma v_w^2\bigr),\\
 \dfrac{dC_d}{dt} &=& \beta C v_p^2 \Bigl(1-\dfrac{C_d}{C_{d,max}}\Bigr).\notag
\end{eqnarray}
Here $D_u$ is the apparent diameter in the upstream direction, $D_s$ is the diameter in the direction perpendicular to fluid flow, while $D_d$ is the downstream diameter. Schematic of the eroded perforation, according to the model, is shown in Fig.~\ref{fig6}.
As can be seen from~(\ref{newerosion}), the erosion in the downstream direction is also affected by the term that is proportional to $\gamma$, which is a dimensionless parameter, as well as the squared fluid velocity in the wellbore $v_w$. This term causes heel bias of the perforation erosion because the fluid velocity in the wellbore is higher at the heel of the stage and is lower at the toe of the stage. In addition, the shape of the eroded perforation qualitatively agrees with the shapes observed in the field~\cite{Cramer2020}.

Similarly to the values of $\alpha$ and $\beta$, the parameter $\gamma$ can vary from one field case to another. The default values in StageOpt are taken from~(\ref{ouralpha}), while $\gamma=100$. In order to match field observations, multipliers for $\alpha$ and $\gamma$ are introduced in StageOpt, while the value of $\beta$ is less crucial and therefore its multiplier is set to be equal to that for $\alpha$. Also note that it is important to have an elongated perforation shape in the model because the ability of proppant to turn into perforation depends on the perforation size along the fluid flow direction, i.e. $(D_u+D_d)/2$. At the same time, perforation friction depends on the total area of the perforation. Therefore, it is useful to introduce the equivalent diameter as
\begin{equation}\label{Deq}
D_{eq} = \sqrt{\dfrac{4A}{\pi}},\qquad A = \dfrac{\pi}{8}D_s(D_u+D_d).
\end{equation}
In this case, the equation for perforation friction still holds if the equivalent diameter is used in it to calculate perforation area. Also, the circumferential and axial diameters, that are often reported in field data, are simply:
\begin{equation}\label{Dcircax}
D_{circ} = D_s,\qquad D_{ax} = \dfrac{1}{2}(D_u+D_d).
\end{equation}
Note that in the model $D_s=D_u$, in which case there are only two independent diameters. Thus, the erosion model~(\ref{newerosion}) can be rewritten in terms of the circumferential and axial diameters as
\begin{eqnarray}\label{newerosion2}
\dfrac{dD_{circ}}{dt}  &=& \alpha C v_p^2,\notag\\
\dfrac{dD_{ax}}{dt}  &=& \alpha C \bigl(v_p^2+\tfrac{1}{2}\gamma v_w^2\bigr),\\
 \dfrac{dC_d}{dt} &=& \beta C v_p^2 \Bigl(1-\dfrac{C_d}{C_{d,max}}\Bigr).\notag
\end{eqnarray}
In this case, perforation area is $A=\pi D_{circ}  D_{ax}/4$ and the equivalent diameter is $D_{eq}=\sqrt{D_{circ}  D_{ax}}$.

To summarize, StageOpt uses the modified erosion model~(\ref{newerosion}) with the default values for $\alpha$ and $\beta$ from~(\ref{ouralpha}), while $\gamma=100$. To calibrate the model, erosion multipliers for $\alpha$ and $\gamma$ are introduced in StageOpt, while $\beta$ uses the same multiplier as $\alpha$.

\section{Optimization and uncertainty}

StageOpt allows users to run the code ``as is'', or to do optimization. Two types of optimizations are implemented, see also~\cite{Donts2023a} and~\cite{Donts2023b} for details. The first one is more theoretical than practical and allows users to optimize phasing of each individual perforation. This case is applicable when the injection schedule consists of a single line and there is no perforation erosion. If the injection schedule is more complex, then ``average'' injection schedule is considered. If erosion is included, then optimization is performed assuming no erosion, while the final result is reported with erosion. This option allows to investigate the most optimal, albeit arguably not practical case, when phasing of each individual perforation can be changed.

A practical optimization is also included. It is assumed that all perforations have the same orientation, but the value of this phasing is unknown. In this case, StageOpt runs a series of calculations for different phasings. The case with the most uniform proppant distribution is selected as optimal. In this optimization, both erosion and full injection schedule are considered.

Proppant and slurry uniformity are characterized by uniformity indices. The index can be calculated at the perforation level or at a cluster level. They are defined for the proppant and slurry as
\begin{equation}\label{uniformityindices}
U_p = 1-\dfrac{\sqrt{\langle (m_p-\langle m_p\rangle)^2\rangle}}{M\langle m_p\rangle},\qquad U_s = 1-\dfrac{\sqrt{\langle (V_s-\langle V_s\rangle)^2\rangle}}{M\langle V_s\rangle}.
\end{equation}
Here $\langle \cdot \rangle$ represents the average operator, $m_p$ is total proppant mass, while $V_s$ is total slurry volume. The average is taken over perforations or over clusters depending on the type of uniformity index. The value of the multiplier $M=1$ for regular uniformity index, while it is taken as $M=\sqrt{N\!-\!1}$ for the normalized uniformity index, where $N$ is either the number of perforations or the number of clusters, see~\cite{Wehunt2020}. The normalized uniformity index is introduced to ensure that its value is between 0 and 1, while the regular uniformity index can be negative.

Field scale operations inevitably lead to uncertainty of some of the parameters. Several types of uncertainty are considered in StageOpt: uncertainty of phasing at the stage level, phasing uncertainty at the perforation level, and perforation diameter uncertainty. Note that there is also tensile strength uncertainty and probabilistic approach is used for perforation plugging, which both introduce uncertainty in the number of active perforations. The stage level phasing uncertainty represents the variation of phasing from stage to stage, but not within the stage. For instance, if $0^\circ$ design phasing is specified and the stage level uncertainty is $10^\circ$, then all perforations within the stage are assigned a phasing value between $-10^\circ$ and $10^\circ$. At the same time, the perforation level uncertainty assigns an uncertainty to each individual perforation independently. When uncertainty is evaluated, the number of Monte Carlo draws needs to be specified. In this case, StageOpt runs a series of simulations with different random realizations and reports statistics for all the cases.

\section{Perforation friction pressure drop}

StageOpt outputs several types of perforation friction pressure drops. The two primary categories are ``theoretical'' and ``actual'' pressure drops. In the ``theoretical'', it is assumed that the flow rate in each perforation is the same. Its value is calculated by dividing the maximum injection rate in the schedule by the number of perforations. For instance, if the maximum injection rate is 90~bpm and there are 30 perforations total, then the calculations are performed assuming 3~bpm per perforation, no matter what the perforation diameters are, what stress shadow values are, etc. In contrast, ``actual'' perforation pressure drop is calculated by using the maximum injection rate and distributing it between the perforation shots accounting for the variability of diameters, near-wellbore friction, stress shadow, etc. Further, the ``theoretical'' perforation friction drop is subdivided into three cases: (a) design pre-frac, (b) true pre-frac, and (c) post-frac. In the design pre-frac, the design perforation diameter (the one specified in the table) is used in calculations. In the true pre-frac, actual initial perforation diameter is used. This diameter accounts for the diameter uncertainty, as well as the variation with respect to phasing or gun clearance. The result is reported in terms of the mean and standard deviation of the resultant perforation friction, which is calculated from all Monte Carlo draws. The post-frac case uses final perforation diameters after the stage is completed and the result is again reported in terms of the mean and standard deviation that are calculated from all Monte Carlo draws. The ``actual'' perforation friction drop has only one case - post-frac, in which the final perforation diameter is used in calculations, and the results are again reported in terms of the mean and standard deviation that are calculated from all Monte Carlo draws. To summarize, StageOpt outputs four different types of perforation friction values: 1) theoretical design pre-frac (uniform rate per shot, design shot diameter), 2) theoretical true pre-frac (uniform rate per shot, actual initial diameter), 3) theoretical post-frac (uniform rate per shot, actual final diameter), and 4) actual post-frac (actual rate per shot, actual final diameter).

\section{Correction for the level of suspension}

Upon comparing StageOpt results with field data, it became apparent that in certain cases particles are suspended better than predicted by StageOpt. There could be numerous reasons for it. One could be a fundamental reason, because the calibration of suspension flow is performed for relatively big particles~\cite{Dont2023c}, while much smaller proppants are often used. As a result, the extrapolation of the calibration may not be sufficiently accurate to predict the behavior for small particles. The second reason is fluid rheology. While the calibration is performed for water, typical hydraulic fracturing fluids often have additives, such as friction reducers, which in turn can significantly alter behavior of the fluid. The way particles are suspended in a fully turbulent flow with such complex rheology fluid is largely unknown. In order to model the difference between the calibrated model and a realistic situation, suspension calibration parameter $S$ is introduced. With the reference to~(\ref{Gdef}), the dimensionless gravity is calculated as
\begin{equation}\label{Gdef2}
 G = \dfrac{8\phi_m(\rho_p\!-\!\rho_f)g \cos(\theta_w)d_w}{S f_D\rho_f v_w^2},\qquad  t_0 = \dfrac{9\mu_a d_w S}{2(\rho_p\!-\!\rho_f)g\cos(\theta_w) a^2},
\end{equation}
where $t_0$ is the characteristic settling time, which is also affected by the parameter $S$. In the latter equation $\mu_a$ is apparent viscosity and $a$ is proppant radius. As can be seen from this modification, the parameter $S$ effectively reduces the density contrast and thus alters the ability of particles to be suspended. If $S>1$, then particles are suspended better than the prediction of the original calibration. 

\section{Correction for inline perforations}

An interesting phenomenon was observed in CFD simulations in~\cite{Liu2021,Wang2022}. It was found that if perforations are located inline, i.e. they all have the same phasing within the cluster, then the amount of proppant that enters each following perforation increases. This is also supported by field observations of erosion patterns, whereby the level of erosion increases for downstream inline perforations within the cluster. The mechanism for this phenomenon is the following. Some particles are unable to make the turn into the first perforation and miss it. But then, they end up increasing the concentration locally at the given perforation azimuth. If the next perforation has the same orientation, then it receives more proppant because of the elevated concentration. This process repeats for the downstream perforations. As was mentioned above, there is a field evidence for this phenomenon and therefore we need to have an ability to model it in StageOpt.

\begin{figure}
\centering\includegraphics[width=0.5\linewidth]{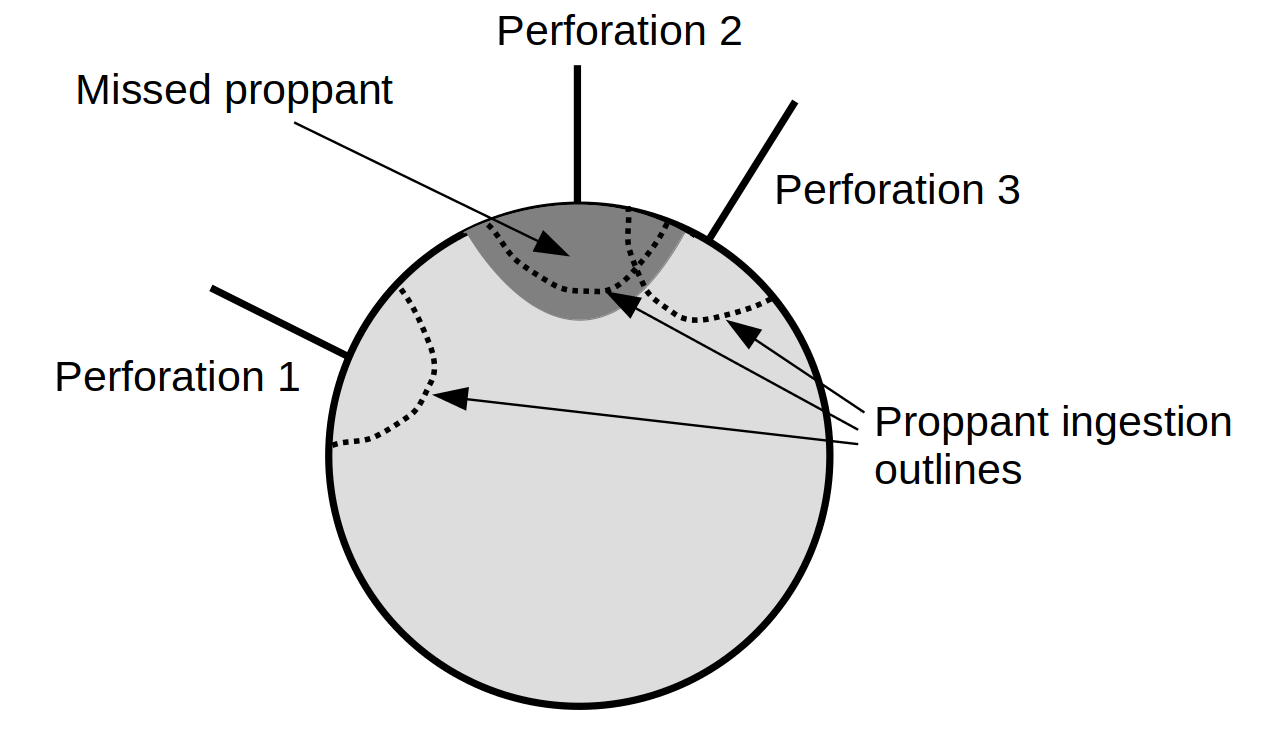}
\caption{Illustration for the inline correction.}
\label{fig7}
\end{figure}

In order to model the effect of missed proppant, the following assumptions are made. First, the missed proppant is located in the vicinity of the perforation that it attempted to flow into. The size of this zone is taken somewhat larger than the original ingestion zone. Second, particle concentration within this zone is assumed to be uniform. Fig.~\ref{fig7} illustrates location of the missed proppant by the dark gray zone, assuming that it tried to enter the perforation oriented upwards or at zero degrees. Note that the light gray color indicates the ``base'' distribution of proppant that is assumed to be uniform for the purpose of this illustration. The amount of missed proppant is multiplied by a calibration parameter called ``inline correction multiplier''. The reason for introducing such a parameter is the following. More particles than just the missed ones are ``attracted'' to the perforation. Even those that were not attempting to enter the perforation are still moved in the direction towards the perforation and thus increase the total amount of proppant that is displaced towards the direction of the perforation. 

Let $\eta$ be the turning efficiency. In other words, this is the fraction of particles that made it through the perforation. In this case, the particle concentration change in the ``missed'' zone is calculated as
\begin{equation}\label{deltaphi}
\Delta \phi = I_c\dfrac{1-\eta}{\eta}\dfrac{q^p_p}{q_sA_m}.
\end{equation}
Here $I_c$ is the inline correction multiplier, $q^p_p$ is proppant flow rate though perforation, $q_s$ is the slurry flow rate in the wellbore after the perforation, and $A_m$ is the area of occupied by the ``missed'' proppant. Concentration of particles in such zones decreases exponentially with time with the time scale defined as
\begin{equation}\label{tinline}
 t_{inline} = \dfrac{9\mu_a d_w^2}{2a^2\rho_f v_w^2}.
\end{equation}

The ``missed proppant'' is defined for every perforation knowing its orientation and the amount missed particles. As was mentioned before, such zones containing the elevated concentration of particles dissipate with time. Practically speaking, they are important within the cluster and the effect fades away once the flow reaches the next cluster. With the reference to Fig.~\ref{fig7}, location of the missed proppant allows us to calculate fraction of this missed proppant that enters the next perforation. Three hypothetical examples are shown in Fig.~\ref{fig7}. The case in which the next perforation has a sufficiently different azimuth (perforation 1), when the next perforation is perfectly aligned with the initial perforation oriented upwards (perforation 2), and the situation for which the next perforation is slightly offset from the initial perforation (perforation 3). The ingestion zones or the zones from which proppant flows into these respective perforations are shown by the dashed lines in Fig.~\ref{fig7}. If there is no overlap between the ingestion zone and the missed proppant, then there is no effect (the case of perforation 1). If the next perforation is perfectly aligned with the previous one (the case of perforation 2), then the effect is maximized and all the proppant that missed the initial perforation is able to enter the next hole. If there is a slight offset with the next perforation, such as if there is an uncertainty in phasing, then the overlap area is calculated and the inline effect is reduced accordingly (the case of perforation 3).

This model, while simple, allows us to capture the primary mechanisms that are relevant for the inline perforations and, most importantly, allows to match field observations that are otherwise impossible to capture.

\section{Erosion uncertainty}

Field observations of perforation erosion indicate that the erosion process has very strong uncertainty. In particular, there is a significant variability from stage to stage, despite the fact that completion is practically the same. There can be different reasons to why this is the case. For instance, there can be a local variability in rock formation immediately behind casing. Alternatively, the quality of cement behind casing can be variable, which can promote flow behind casing and elevate the erosion. No matter what the mechanism is, observations suggest that there is a spatial correlation, i.e. zones with high level of erosion tend to be relatively localized within a stage. 

In order to model this phenomenon, we introduce uncertainty in the perforation erosion process by specifying an erosion multiplier. In particular, there are the following three parameters. The first one is the normalized standard deviation at distance zero $\sigma_0$. The second one is the normalized standard deviation at the specified distance called range $\sigma_R$, while the last one is the range itself $R$. In the context of the variogram, nugget is simply $\sigma_0^2$, sill is $\sigma_R^2$, while range is actually the specified range $R$. In order to introduce the spatially correlated uncertainty, multivariate normal distribution is used, for which the covariance matrix is calculated using the nugget, sill, and range mentioned above. Spherical model for the variogram is used, which is defined as
\begin{equation}
\gamma(d/R) = \sigma_0^2+(\sigma_R^2-\sigma_0^2)\Bigl(\dfrac{3d}{2R}-\dfrac{d^3}{2R^3}\Bigr),
\end{equation}
where $d$ is the distance between the perforations and $R$ is the range. Note that when $d/R>1$, then $\gamma(d/R)=\sigma_R^2$.

Once the spatially correlated uncertainty is calculated for each perforation within the stage, then one of the following equations is used to calculate the erosion multiplier
\begin{equation}
\alpha_m = \dfrac{1+|\alpha_u|}{\langle 1+|\alpha_u| \rangle},\qquad \alpha_m = 1+10^{\alpha_u},\qquad \alpha_m = 1+|\alpha_u|.
\end{equation}
Here $\alpha_m$ is the erosion multiplier for the given perforation, while $\alpha_u$ is the corresponding uncertainty that is calculated from the multivariate normal distribution. The first option corresponds to the ``lognormal'' option, the second one to the ``lognormalnotnormalized'', while the last one is the ``normal'' option. The first option is normalized by the average erosion multiplier to ensure that the uncertainty does not increase the overall rate of erosion.

These options can be entered using advanced settings as:
\\
\\
Name:\\
sill\\
Value(s):\\
0.9\\
\\
Name:\\
nugget\\
Value(s):\\
0.3\\
\\
Name:\\
range\\
Value(s):\\
50\\
\\
Name:\\
erosionunceraintydistribution\\
Value(s):\\
Lognormal\\
\\
Recall that nugget is simply $\sigma_0^2$, while sill is $\sigma_R^2$.

\section{History matching workflow with StageOpt}

This section briefly outlines a history matching workflow that is suggested for StageOpt users. A more detailed version with examples can be found in~\cite{Donts2024}. The workflow for optimizing perforation design employs StageOpt simulator and observation of field-scale phenomena, such as downhole imaging that provides cost-effective and high-fidelity assessment of the magnitude and location of perforation erosion within a stage~\cite{Cramer2020,Robinson2020}. The data can include the equivalent diameter, as well as axial and circumferential diameters. The recommended workflow is:

\begin{itemize}

\item Identify optimization objective and outline the perforation design parameters that are considered for modification as part of an optimization workflow. For instance, this can be perforation phasing, diameter, number of shots per cluster, stage length, and cluster spacing. 

\item Gather datasets for calibration, for which downhole imaging of post-job perforation diameter is available. Ideally, each well needs to be analyzed independently, but several wells can be grouped if they lie in the same formation and have similar casing. If the downhole imaging data is not available, then default calibration parameters can be used as a starting point. 

\item Evaluate statistics from field observations. This helps assigning the magnitudes of uncertainties, such as for phasing and perforation diameter. Also, this step provides the primary history matching quantity of interest, which is the degree of erosion for each shot as well as its variability from stage to stage. It is important to add as many stages as possible to the analysis to increase the level of confidence in the observation data.
        
\item History match the model to the observed erosion data. This primarily entails varying the erosion multipliers, stress shadow (internal and external), as well as perforation tensile strength and plugging (if the corresponding data is available).
 
\item Perforation design optimization. At this step, vary the optimization variables to identify the design that maximizes the uniformity index of proppant and/or slurry or other specified objective. StageOpt reports the uniformity index defined on a ‘per cluster’ basis and a ‘per shot’ basis. Typically, the ‘per cluster’ uniformity is the desired optimization objective, since it is often assumed that the slurry exiting the perforations within the same cluster mixes together outside the casing.
        
\end{itemize}



\end{document}